\documentclass{elsart}

\usepackage{graphics}

\usepackage{amssymb}
\newcommand{\mbf}[1]{\mbox{\boldmath $#1$}}
\newcommand{\beq}{\begin{equation}}   
\newcommand{\eeq}{\end{equation}}      
\newcommand{\beqn}{\begin{eqnarray}}
\newcommand{\eeqn}{\end{eqnarray}}
\newcommand{\ba}{\bar{k}_1}
\newcommand{\bb}{\bar{k}_2}
\newcommand{\bh}{\bar{h}}

\newcommand{\kk}{\mbf{k}}
\begin{document}

\begin{frontmatter}



\title{Perturbative QCD Odderon and Diffractive $\eta_c$ Production}


\author[email]{G. P. Vacca}

\address{ II. Institut f\"ur Theoretische Physik,
Universit\"at Hamburg, Luruper Chaussee 149, D-22761 Hamburg, Germany}

\address[email]{E-mail: vacca@bo.infn.it}

\begin{abstract}
The Odderon state in perturbative QCD is briefly reviewed.
Recent calculations devoted to estimate the diffrative
$\eta_c$ production at high energies from the leading known Odderon states are
discussed.
\end{abstract}

\end{frontmatter}

\section{Introduction}
The Odderon \cite{Lukaszuk:1973nt} is the partner of the Pomeron, but
odd under parity $P$ and charge conjugation $C$ (like the photon).
In general it is related to the possibility that the real part of a
scattering amplitude increases with energy as fast as the imaginary part.
The general analytic properties of the scattering amplitudes at high
energies are conveniently studied in the complex angular momentum plane $j$:
in the even (under crossing) amplitude a singularity is associated to
the Pomeron and gives a mostly imaginary contribution,
while in the odd case one has a mostly real contribution which is associated
to the Odderon. The position of the singularity, near $j=1$, is also called
intercept and it is related to the asymptotic behaviour of the cross section.

It is remarkable that QCD, the very successful theory for the strong
interaction, predicts the existence of the Odderon.
This is related to the fact that the internal gauge symmetry group of QCD
has rank greater than one, and therefore allows to construct from three
gluons a $C$-odd state which can be associated to the Odderon.
Infact, on considering the $SU(3)_C$ gauge group associated
to the gluon field $A_\mu=\sum_a A_\mu^a t_a$, and using the transformation
property $A_\mu \to -A_\mu^T$ under charge conjugation, one can see that
two possible independent invariants can be constructed by three gluon fields,
$Tr( [A_1,A_2]A_3 )$ and $Tr( \{A_1,A_2\}A_3 )$,
which are respectively even and odd under charge conjugation.
Therefore the Odderon is related to the composite operator
$O_{\alpha\beta\gamma}=d_{abc}A_\alpha^a A_\beta^b A_\gamma^c$.

The challenge to understand the QCD dynamics present in the scattering processes
involving the Odderon quantum numbers is still open both on theoretical and
experimental sides. Till now no evidence of the Odderon has been found showing
that the cross sections involved are small.
The experimental search can go in two directions.
One can look at the comparison of the total and/or elastic
cross sections for direct and cross-symmetric scattering processes,
like for example in the case of $pp$ and $p\bar{p}$ scattering.
Another class of scattering processes, where the Odderon contributes, is when one or two
of the incoming scattering particles, of definite C-parity, goes into a state of
opposite C-parity under scattering.
A rapidity gap, which allows to separate the outgoing scattering states, is required.
For example, one can look at the reaction
\begin{equation}
\gamma \, (\gamma^*) + p \to PS \, (T) + p \, (X_p), \nonumber
\end{equation}  
where a photon scatters a proton and a pseudoscalar or a tensor meson is produced in
the photon fragmentation region, well separated in rapidity from the proton or its debris
($X_p$). This process has been started to be analyzed at HERA \cite{Schafer:1992pq}. 
Recent non perturbative studies \cite{Heidelberg},
based on some specific dynamical QCD assumptions, and
carried on for the production of light mesons ($\pi^0$, $f_2$),
have been clearly excluded by the recent analysis at HERA from the H1 collaboration
\cite{H1},
where in both cases a much lower limit to the cross section has been posed.

On the other hand perturbative analysis have been performed in the study of
$\eta_c$ production in DIS with an Odderon made by three simply uncorrelated gluons
\cite{Czyzewski:1997bv,Engel:1998cg}
and later by considering the resummed QCD interaction in LLA \cite{Bartels:2001hw},
where leading gluon correlation phenomena are taken into account.
This analysis will be shortly reviewed here.
\section{Perturbative QCD Odderon in LLA}
In the high energy limit we consider a scattering process where all the leading
log contributions in the center of mass energy are resummed, i.e.
those of the order $(\alpha_s \ln s)^n$.
The so called $k_T$ factorization does apply and the amplitude is given by
$A(s,t) = c \langle \Phi_1|G|\Phi_2\rangle$,
where $c$ is a convenient normalization factor, $\Phi_i$ are the impact factors
and $G$ is the Green function, describing an effective evolution in rapidity.
Here a bra-ket notation is used for the integration
on the transverse plane. Normally, coordinate or momentum representations are chosen.

For the Odderon case,
at lowest order, without resummation, and provided the strong coupling
$\alpha_s$ is small, one has a simple three uncorrelated gluon exchange,
i.e. the Green function $G_3$, which is
convoluted with the impact factors, is constructed, simply with 3 gluon propagators.
Therefore, in momentum representation
$G_3^{(LO)}=\delta^{(2)}(\mbf{k}_1-\mbf{k}'_1)\delta^{(2)}(\mbf{k}_2-\mbf{k}'_2)
1/\mbf{k}_1^2\mbf{k}_2^2\mbf{k}_3^2$.

In the high energy limit, when all other physical invariants are much smaller,
the LLA resummation can be performed.
The same resummation for the two gluon exchange has lead to the BFKL \cite{BFKL}
equation. The kernel of this integral equation for the 2-gluon Green function,
in the colour singlet state, describes the perturbative QCD Pomeron in LLA.
The same equation in the colour octet state has a simple eigenstate which corresponds
to the reggeized gluon, a composed object at high energies.
This fact is seen as a self consistency requirement and it is called bootstrap.
In NLA \cite{Fadin:1998py}, where one is resumming also the contribution of order
$\alpha_s^n (\ln s)^{n-1}$, all the same concepts, reggeization included
\cite{bootstrap}, do apply.

The general kernel for the $n$-gluon integral equation for the Green function in LLA is
given by the BKP equation \cite{BKP}. In the large $N_c$ limit and for finite $N_c$ when
$n=3$, it possesses remarkable symmetry properties:
discrete cyclic symmetry, holomorphic separability, conformal invariance, integrability,
duality \cite{symmetries} and also a relation between solutions with different $n$
exists \cite{Vacca:2000bk}, which is a direct consequence of the gluon reggeization.

The Odderon states in LLA must be symmetric eigenstates of the operator
$K_3=1/2(K_{12}+K_{23}+K_{31})$ constructed with the BFKL kernel $K_{ij}$ for two
reggeized gluons in a singlet state.
Using the conformal invariance and integrability properties a set of eigenstates
has been found \cite{Janik:1999xj}, which have a maximal intercept below one. 

Using the gluon reggeization property (bootstrap) a new set of solutions was later
found   \cite{Bartels:2000yt}, characterized by intercept up to one,
therefore dominant at high energies. Moreover for the particular impact factor
which couples a photon and an $\eta_c$ to the Odderon the LLA calculation has shown
that this second set of solution is relevant while the previous one decouples.

These leading Odderon states $E_3^{(\nu,n)}$, in momentum representation, are given by
\begin{equation}
\mbf{k}_1^2 \mbf{k}_2^2 \mbf{k}_3^2 \, E_3^{(\nu,n)}(\mbf{k}_1,\mbf{k}_2,\mbf{k}_3)=
c(\nu,n) \sum_{(123)} (\mbf{k}_1+\mbf{k}_2)^2 \mbf{k}_3^2 \, 
E^{(\nu,n)}(\mbf{k}_1+\mbf{k}_2,\mbf{k}_3),
\label{oddwave}
\end{equation}
where $c(n,\nu)=\frac{1}{(2\pi)^{3/2}}
\sqrt{\frac{g_s^2 N_c}{-3\chi(\nu,n)}}$
is a normalization factor, $E$ is a BFKL Pomeron eigenstate
and the conformal spin $n$ is odd.
The sum is taken over the cyclic permutations.
In the expression above no colour wave function is explicit.
The full Green function is constructed summing over all such states in the following way,
in the momentum representation,
\beqn
\hspace*{-0.8cm} G_3(y|\mbf{k}_1,\mbf{k}_2,\mbf{k}_3|\mbf{k}'_1,\mbf{k}'_2,\mbf{k}'_3)&=&
\sum_{{\rm odd}\ n}\int_{-\infty}^{+\infty} d\nu
\frac{(2\pi)^2(\nu^2+n^2/4)}{[\nu^2+(n-1)^2/4][\nu^2+(n+1)^2/4]} \times \nonumber \\
&& e^{y\, \chi(\nu,n)} E_3^{(\nu,n)}(\mbf{k}_1,\mbf{k}_2,\mbf{k}_3)
{E_3^{(\nu,n)}}^*(\mbf{k}'_1,\mbf{k}'_2,\mbf{k}'_3) \, .
\label{greenf}
\eeqn
In the high energy limit the asymptotic behaviour can be studied for conformal spin
$n=\pm 1$ and performing the saddle point integration around $\nu=0$.

Since we have found convenient to work in the momentum representation to make a
close comparison between the LLA Odderon contribution and the simple 3 gluon exchange,
we need the pomeron BFKL function $E$ in such a representation.
The Fourier transform was performed in \cite{Bartels:2001hw} and is given by
\beq
\tilde{E}_{h\bh}(\kk_1,\kk_2)= \tilde{E}_{h\bh}^{A}(\kk_1,\kk_2)+
 \tilde{E}_{h\bh}^{\delta}(\kk_1,\kk_2),
\label{pom_mom}
\eeq
where $h=(1+n)/2+i\nu$ and $\bar{h}=(1-n)/2+i\nu$ are called conformal weights.
The analytic term ( also standard complex notation is used) is given by
\beq
\hspace*{-0.4cm}\tilde{E}_{h\bh}^{A}(\mbf{k}_1,\mbf{k}_2)=
\frac{(-i)^n}{(4\pi)^2}h\bh \Gamma(2-h)\Gamma(2-\bh)
\Bigg[X(\mbf{k}_1,\mbf{k}_2)+
(-1)^nX(\mbf{k}_2,\mbf{k}_1)\Bigg],
\label{pom_an}
\eeq
where
\beq
\hspace*{-0.9cm} X(\mbf{k}_1,\mbf{k}_2)\hspace*{-0.1cm}=\hspace*{-0.1cm}
\left(\frac{k_1}{2}\right)^{\bh-2}\hspace*{-0.2cm}
\left(\frac{\bb}{2}\right)^{h-2}\hspace*{-0.3cm}
F\hspace*{-0.1cm}\left(1-h,2-h;2;-\frac{\ba}{\bb}\right)
\hspace*{-0.1cm}
F\hspace*{-0.1cm}\left(1-\bh,2-\bh;2;-\frac{k_2}{k_1}\right)\hspace*{-0.1cm}.
\label{fullX}
\eeq
The $\delta$-distribution term is
\beq
\tilde{E}^{\delta}_{h\bh}(\mbf{k}_1,\mbf{k}_2)=
\Bigl[ \delta^{(2)}(\mbf{k}_1) +(-1)^n \delta^{(2)}(\mbf{k}_2) \Bigr]
\frac{i^n}{2\pi} 2^{1-h-\bh} \frac{\Gamma(1-\bh)}{\Gamma(h)} 
q^{\bh-1} q^{*\, h-1 }, 
\label{pom_delta}
\eeq
where $\mbf{q}= \mbf{k}_1+\mbf{k}_2$.
All the expression needed in the case $n=\pm 1$ and $\nu \to 0$ are however strongly
simplified \cite{Bartels:2001hw}.
\section{The process $\gamma^* p \to \eta_c p$.}
We shortly review the calculation of the amplitude
\beq
A^i(s,t) = \frac{s}{32}\frac{1}{16}\frac{N_c^2-4}{N_c}\frac{N_c^2-1}{3!}
\frac{1}{(2\pi)^8} \langle \Phi^i_{\gamma}|G_3|\Phi_p\rangle.
\label{ampli}
\eeq
The impact factors are the ones presented in \cite{Czyzewski:1997bv,Engel:1998cg}
in order to have a direct comparison of the resummation effects.
Therefore we use the following photon impact factor
\beq
\Phi_{\gamma}^i=b \, \epsilon_{ij} \frac{q_j}{\mbf{q}^2}
\left( \sum_{(123)} 
\frac{(\mbf{k}_1 + \mbf{k}_2 - \mbf{k}_3) \cdot \mbf{q}}{Q^2+4m_c^2 + 
(\mbf{k}_1 + \mbf{k}_2 - \mbf{k}_3)^2}
-\frac{\mbf{q}^2}{Q^2+4m_c^2 + \mbf{q}^2} \right) \, , 
\label{impactgamma}
\eeq
where $b$ is a normalization factor \cite{Czyzewski:1997bv,Engel:1998cg}.
It is calculated
in perturbation theory, thanks to the presence of the charm scale, with some simplifying
assumptions. The transverse polarization case is considered since in particular the real
photon case is giving the dominant contribution.
Its form was discussed also in \cite{Bartels:2000yt}. The longitudinal polarization play
no important role for our purposes.
The proton impact factor cannot be clearly calculated in perturbation theory and only
some reasonable ansatz can be given. In particular we have
\beq
\Phi_p=d\Big[F(\mbf{q},0,0)-\sum_{i=1}^3F(\mbf{k}_i,
\mbf{q}-\mbf{k}_i,0)+2F(\mbf{k}_1,\mbf{k}_2,\mbf{k}_3)\Big]\, ,
\label{proton_if}
\eeq
with
\beq
F(\mbf{k}_1,\mbf{k}_2,\mbf{k}_3)=
\frac{2a^2}{2a^2+(\mbf{k}_1-\mbf{k}_2)^2+
(\mbf{k}_2-\mbf{k}_3)^2+(\mbf{k}_3-\mbf{k}_1)^2}\, ,
\label{Fp}
\eeq
with $d$ and $a$ some parameters for the normalization and the internal proton scale
\cite{Czyzewski:1997bv,Engel:1998cg}.

As anticipated, we evaluate the amplitude (\ref{ampli}) in the saddle point approximation
for $n=\pm 1$ conformal spin. The integral (scalar product) of the Odderon states
with the photon impact factor can be computed analytically in this limit while
the proton side has instead to be treated numerically \cite{Bartels:2001hw} and
some care is needed to assure the cancellations of the infrared singularities.

After having performed the saddle point integration we have obtained an approximation
for the amplitude which we can use to calculate the differential cross section,
squaring it and averaging over the photon polarizations: 
\beqn
\hspace*{-0.8cm}\frac{d\sigma}{dt}(\gamma(\gamma^*)+p\rightarrow\eta_c+p)&=&
\frac{1}{16\pi s^2}\ \frac{1}{2}\sum_{i=1}^2|A^i|^2=
\frac{2^4 \cdot 5^2 }{3^7}\frac{1}{(2\pi)^8}
\frac{\alpha_{em} \alpha_s^2 b_0^2}{\zeta(3)y}\frac{1}{|t|}
\times \nonumber \\
&&\frac{m_{\eta_c}^2}{(Q^2+4 m_c^2)2 a^2}
|V^{(0,\pm 1)}_\gamma(\frac{t}{M^2})|^2|V^{(0,\pm 1)}_p(\frac{t}{2a^2})|^2.
\label{cross1}
\eeqn

In the above expression one has \cite{Bartels:2001hw}
\beq
V_{\gamma}^{(0,\pm 1)}\left(\frac{t}{M^2}\right) = 
\frac{\sqrt{|t|/M^2}}{1+|t|/M^2} \, , \nonumber
\eeq
where $M=\sqrt{Q^2+4m_c^2}$,
while $V_p^{(0,\pm 1)}$, which has been evaluated numerically, is shown in Fig. 1.

\begin{figure}
\label{Fig1}
\centering
\includegraphics*{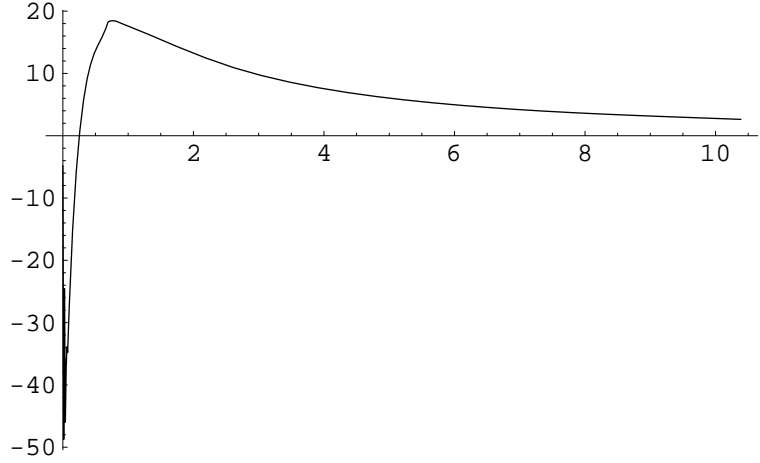}
\caption{Numerical results for the coupling of the Odderon to the proton,
as a function of the scaled variable $x=|t|/2a^2$.}
\end{figure}

It is interesting to see that in the high energy limit the effective
coupling of the Odderon function with the proton impact factor is changing sign with $t$,
and in particular there is an oscillation.
When other non leading states of the Odderon basis are contributing,
this fact could disappear.  Moreover the photon-$\eta_c$-Odderon coupling does not
presents such behaviour.

This fact leads to the presence of a zero in the differential cross section which should,
more generally, present therefore a dip at high energies.

If one looks at the form of the pomeron eigenstates,
one can explain this looking at their $t$
dependence. This is a feature which appears in a non forward analysis and
can also help to understand if an asymptotic regime, where saddle point
integration can be performed, is reached.  
On the other side the effect for the Odderon exchange is more complex than in the Pomeron
exchange, due to the symmetry properties of the solution and the richer structure of
the three gluon impact factors.

The differential cross section is presented in Fig. 2 for the two cases $Q^2=0$  and 25 
GeV$^2$ and $\sqrt{s}\approx 300$ GeV. 
We have taken $\alpha_s$ at the scale $m_c^2 +Q^2$
(in \cite{Czyzewski:1997bv} at $Q^2=0$ the scale was $m_c^2$).
At $t=0$ the cross section vanishes, as in
the case of a simple three gluon exchange.
For the integrated cross sections we find $50$ pb and $1.3$ pb at
$Q^2=0$ and  25 GeV$^2$, respectively.
Compared to the value in \cite{Czyzewski:1997bv} we found a total
cross section an order of magnitude larger.
Note that  compared to the simple
three gluon exchange,  we have a (weak) logarithmic suppression with
energy. So the obtained enhancement effect is totally due to the coupling 
of the Odderon wave function to the impact factors.

There are, in conclusion, many new interesting features appearing after LLA
resummation for processes related to the Odderon exchange.
There are many uncertainties, also due to the
non perturbative ansatz for the proton impact factors, but some of them may could be
fixed making comparisons with other processes and some of the qualitative
features we find, if observed at very high energies, could strongly support
the perturbative Odderon structure till now understood.
By the way the predicted cross sections are nowdays too small
for an experimental check with available data. 

\begin{figure}
\label{Fig2}
\centering
\includegraphics*{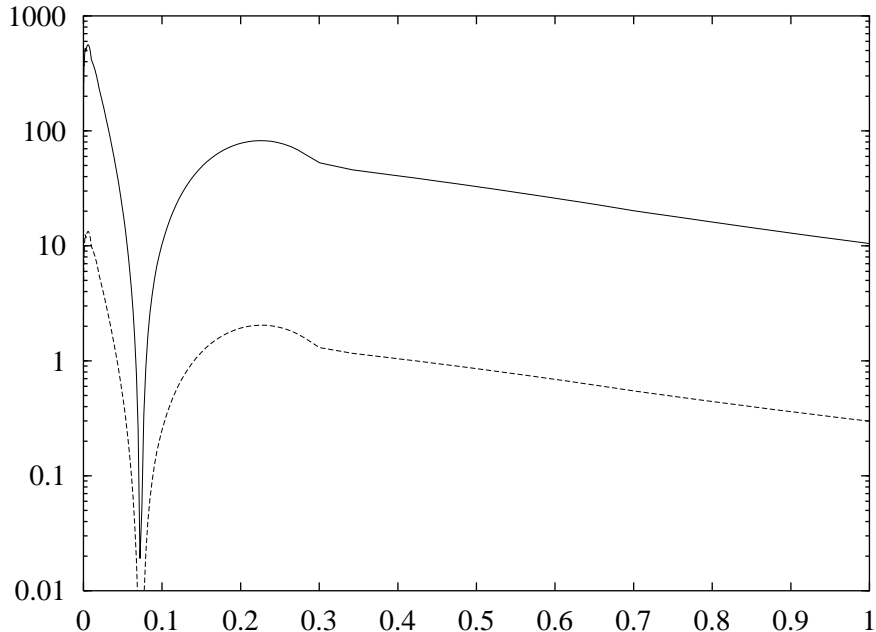}
\caption{The differential cross sections (in pb $/$ GeV$^2$). 
The upper curve refers to $Q^2=0$, the lower one to $Q^2=25$ GeV$^2$.}
\end{figure} 
\noindent
{\bf Acknowledgments:}

\noindent
This talk is mainly based on the results obtained in reference
\cite{Bartels:2001hw}. Many thanks go to the organizers of this very nice workshop. 



\end{document}